\documentclass[shawpacs,aps,graphicx,twocolumn]{revtex4}
\usepackage{amsmath}
\usepackage{graphicx}
\begin{document}


\title{Error-heralded generation and self-assisted complete analysis of two-photon hyperentangled Bell states through single-sided quantum-dot-cavity systems}

\author{Yan-yan Zheng, Lei-xia Liang, and Mei Zhang$^{a,}$\footnote{email:
zhangmei@bnu.edu.cn}}

\address{$^a$ Department of Physics, Applied Optics Beijing Area Major Laboratory,
Beijing Normal University, Beijing 100875, China}

\date{\today }

\begin{abstract}
Hyperentangled Bell-state analysis (HBSA) is critical for high-capacity quantum communication. Based on a recent proposal by Wang \emph{et al.} [Opt. Express 24 28444--28458 (2016)].
We design two separate schemes for error-heralded deterministic generation and self-assisted complete HBSA of two-photon entangled in both polarization and spatial-mode degrees of freedom. Different from previous programs, we firstly proposed an error-heralded block with a singly charged quantum dot inside a single-sided optical microcavity, with which errors due to imperfect interactions between photons and quantum dot systems can be heralded. Thanks to the error-heralded block, the fidelity of the two schemes for hyperentangled Bell-state generation and complete HBSA can reach unit one. Besides, hyperentanglement makes it possible to analyze the polarization state assisted by the measured spatial-mode state. The self-assisted way of the HBSA greatly simplifies the analysis process and largely relaxes the requirements on nonlinearities. Therefore, the schemes hold the promise to implement more easily in experiments, taking a step closer to the long-distance high-capacity quantum communication.
\end{abstract}


\maketitle

\section{introduction}
Quantum entanglement is a critical quantum resource in quantum information processing  \cite{shu}, and has attracted much attention in recent years due to its numerous applications \cite{Bennett,Ekert,LXHQKD,Wiesner,Hillery,Long,Deng,WangCQSDC,QSDC3,QSDC4,QSDC5}. In quantum communication, information is encoded in different quantum states, for example the four Bell states of a two-photon entangled system. In order to read the quantum information, we must completely distinguish the different quantum states, Bell-state analysis (BSA) is an essential technology. In 1999, two schemes for analyzing the photon polarization degrees of Bell state using only linear optics were proposed by Vaidman \emph{et al.} \cite{Vaidman} and L$\ddot{u}$tkenhaus \emph{et al.} \cite{Ltkenhaus}, but the maximum success rate of their schemes is only 50$\%$ theoretically and experimentally. Some other proposals \cite{Kwiat,Walborn,Schuck,Barbieri} show that by introducing entanglement of other degrees of freedom, four Bell states of a single degree of freedom can be completely distinguished with only linear optical elements. 

In order to further improve the quantum channel capacity, beat the limit of linear photonic superdense coding \cite{Barreiro}, we can also encode the quantum information in more than one degrees of freedom. Hyperentanglement \cite{hyper1,hyper2,hyper3,hyper4,hyper5,hyper6,hyper7,hyperentanglementreview}, which means particles are simultaneously entangled in multiple degrees of freedom (DOFs), becomes a key resource in high-capacity quantum communication\cite{HQC2,HQC3,Wei-hyperparallel-computing,HQC5,HEPPWangGY1,HECP1,HECPRenLong,HECPRenDeng,HECPLixh1,HECPLixh2,HECPLixh3}. Currently, the preparation of hyperentangled states in different DOFs, such as polarization-momentum DOFs \cite{hyper2}, polarization-orbital-angular momentum DOFs \cite{hyper4}, and multipath DOFs \cite{hyper5}, are already available. In experiments, the techniques used for creating a single DOF entanglement can be combined to generated hyperentanglement \cite{hyper6}. However, completely analysis of the hyperentangled Bell-state is still a huge challenge in high-capacity quantum information processing. Considering that completely HBSA cannot be accomplished only with linear optical elements \cite{HBSA,HBSA1}, researchers have gradually introduced nonlinear media \cite{Sheng2,Ren,Wang1,Liu1,Li,WangGYHBSA,LiuQ2HBSA,Li3linearanalysis} to assist complete HBSA. In 2010, Sheng \emph{et al.} \cite{Sheng2} firstly presented a way to completely distinguish the 16 hyperentangled Bell states completely with cross-Kerr nonlinearity, and discussed its application in quantum hyperteleportation and hyperentanglement swapping. Subsequently, some other interesting HBSA schemes with single-sided \cite{Ren} or double-sided  \cite{Wang1} quantum-dot (QD) cavity system were proposed. In 2015, two schemes for hyperentangled Bell-state generation (HBSG) and HBSA using nitrogen-vacancy (NV) centers cavity system were proposed by Liu \emph{et al.} \cite{Liu1}, which can realize non-destructive discrimination of 16 two-photon hyperentangled states by using four NV systems. Although the use of nonlinearity can achieve the complete distinction of the hyperentangled Bell state, due to the nonlinear interaction, its fidelity is difficult to reach unit one, and the efficiency can not reach 100$\%$. In 2016, Li \emph{et al.} \cite{Li} proposed a self-assisted complete HBSA scheme using cross-Kerr nonlinearity, which use spatial information to assist polarization state differentiation, the nonlinear interactions were reduced and the fidelity were increased to some extent. Then, Wang \emph{et al.} \cite{WangGYHBSA} gave a way for error-detected HBSG and complete HBSA for photon systems assisted by double-sided QD-cavity systems. In which, the errors caused by nonlinear interactions can be detected, and their fidelities can be
largely improved. In 2016 \cite{LiuQ2HBSA} and 2017 \cite{Li3linearanalysis}, Liu \emph{et al.} and Li \emph{et al.} proposed two solutions respectively, they try to assist HBSA with a third degree of freedom take the place of some nonlinear components. Inspired by these works, we want to propose a new solution that will enable the fidelity to reach unit one and maintain a high efficiency.

In our scheme, an error-heralded block was constructed with QD in a single-sided cavity system and two schemes for deterministic HBSG and complete HBSA were proposed with this block. As the errors caused by the imperfect interaction between photons and QD-cavity systems can be heralded, the two schemes have a high fidelity of unit one. The error-heralded HBSG scheme can be repeated until the deterministic generation of hyperentangled two-photon state was accomplished. By using the measured spatial-mode state to assist the analysis of the polarization state, the complete HBSA scheme works in a self-assisted way and uses only two nonlinear QDs, and it can also realize the function of error avoidance, which greatly reduces the resource loss. Compared to a double-sided cavity system, a single-sided cavity system does not require the balance between the upper and lower sides, and is easier to construct experimentally. The schemes are more useful in high-capacity quantum communication with hyperentanglement in the future.
\section{Interaction between a circularly polarized light and a singly charged QD in a
single-sided microcavity}\label{sec2}
\begin{center}
\begin{figure}[!h]
\includegraphics[width=6 cm]{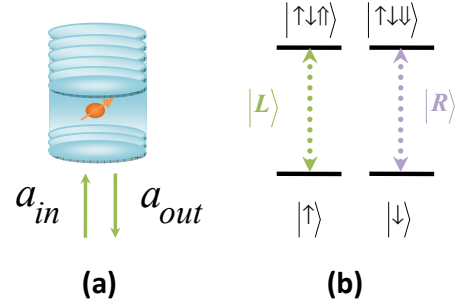} \caption{(a) Schematic
diagram of a singly charged QD inside a single-sided optical
micropillar cavity . (b) The relative energy levels and the optical
transitions of a QD.} \label{figure1}
\end{figure}
\end{center}

Semiconductor quantum dots are a popular solid physical system applied to quantum information research \cite{Wei-hyperparallel-computing,HQC5,Wang1,Hu2,WangCEleEPPECP,Ugate1,WangGYHBSA,LiThereldedrepeater,An,Hu1,Hu3,Ren,HECP1,error-rejectgate1,error-rejectgate2} in recent years because of its long coherence time \cite{Petta,Greilich} and good integration \cite{Atatre,Berezovsky,Press}. Fig.\;\ref{figure1}(a). shows the singly charged quantum dot(e.g., a self-assembled In(Ga)As QD or a GaAs interface QD) embedded in the center of a single-sided microcavity (the top distributed Bragg reflectors are 100\% reflective and the bottom distributed Bragg reflectors are partially reflective).

If an extra electron is injected into the quantum dot, a negatively charged exciton \cite{Warburton}($X^-$, $|\uparrow\downarrow\Uparrow\rangle$ or $|\downarrow\uparrow\Downarrow\rangle$) containing two electrons and a hole will be created with the optical excitation shown in Fig.\;\ref{figure1}(b), where the selection criteria of spin-dependent transition is based on the Pauli's exclusion principle \cite{Hu5}. That is, if the excess electron is in the state $|\uparrow\rangle$, the negatively charged exciton $X^-$ in the state $|\uparrow\downarrow\Uparrow\rangle$ is created by absorbing the left circularly polarized light $|L\rangle$. If the excess electron is in the state $|\downarrow\rangle$, the right circularly polarized light $|R\rangle$ can be absorbed to create the negatively charged exciton $X^-$ in the state $|\downarrow\uparrow\Downarrow\rangle$. Here, $|\Uparrow\rangle$ ($|\Downarrow\rangle$) and $|\uparrow\rangle$ ($|\downarrow\rangle$) represent a hole state and an electron state with the spin projections $|+\frac{3}{2}\rangle$  ($|-\frac{3}{2}\rangle$) and $|+\frac{1}{2}\rangle$ ($|-\frac{1}{2}\rangle$), respectively.

In the interaction picture \cite{Walls}, the Heisenberg equations of cavity field operator $\hat{a}$ and exciton $X^-$ operator $\hat{\sigma}_{-}$  are described as

\begin{eqnarray}                
\begin{split}
\frac{d\hat{a}}{dt}&=-[i(\omega_{c}-\omega)+\frac{\kappa}{2}+\frac{\kappa_{s}}{2}]\hat{a}-g\hat{\sigma}_{-}-\sqrt{\kappa}\hat{a}_{in},\\
\frac{d\hat{\sigma}_{-}}{dt}\!\!&=\!\!-[i(\omega_{X^{-}}-\omega)+\frac{\gamma}{2}]\hat{\sigma}_{-}-g\hat{\sigma}_{z}\hat{a},\\
\hat{a}_{out}&=\hat{a}_{in}+\sqrt{\kappa}\hat{a},
\end{split}
\end{eqnarray}
where $\omega$, $\omega_{c}$ and $\omega_{X^{-}}$ denote the frequencies of the input photon, the columnar microcavity and the $X^{-}$ transition, respectively. $g$ is the coupling strength between the quantum dot and the microcavity. $\gamma/2$ and $\kappa/2$ represent the decay rates of $X^{-}$ and microcavity field mode. $\kappa_{s}/2$ is  the microcavity leakage rate, $\hat{a}_{in}$ and $\hat{a}_{out}$ represent input and output field operators, respectively.

In the weak excitation approximation ($\langle\hat{\sigma}_{z}\rangle=-1$ and $\hat{\sigma}_{z}\hat{a}=-\hat{a}$), the solution of the reflection coefficient \cite{Hu1,An} of the single-sided quantum-dot-cavity system can be obtained as
\begin{small}
\begin{eqnarray}            
r(\omega)=1-\frac{\kappa[i(\omega_{X^{-}}-\omega)+\frac{\gamma}{2}]}{[i(\omega_{X^{-}}-\omega)+
\frac{\gamma}{2}][i(\omega_{c}-\omega)+\frac{\kappa}{2}+\frac{\kappa_{s}}{2}]+g^{2}}.\nonumber\\
\end{eqnarray}
\end{small}
In the cold cavity case ($g=0$), the reflection coefficient reduces to
\begin{eqnarray}         
r_{o}(\omega)=\frac{i(\omega_{c}-\omega)-\frac{\kappa}{2}+\frac{\kappa_{s}}{2}}{i(\omega_{c}-\omega)+\frac{\kappa}{2}+\frac{\kappa_{s}}{2}}.
\end{eqnarray}
Then the interaction of the photon and the single-sided quantum-dot-cavity system can be described by the reflection operator $\hat{r}(\omega)$. Here
\begin{small}
\begin{eqnarray}         
\begin{split}
\hat{r}(\omega)&=|r_{o}(\omega)|e^{i\varphi_{o}}(|R\rangle\langle
R|\otimes|\uparrow\rangle\langle\uparrow|+|L\rangle\langle
L|\otimes|\downarrow\rangle\langle\downarrow|)\\
& +|r_{h}(\omega)|e^{i\varphi_{h}}(|L\rangle\langle
L|\otimes|\uparrow\rangle\langle\uparrow|+|R\rangle\langle
R|\otimes|\downarrow\rangle\langle\downarrow|),
\end{split}
\end{eqnarray}
\end{small}
where $\varphi_{o}=\arg[r_{o}(\omega)]$ and $\varphi_{h}=\arg[r_{h}(\omega)]$ denote the phase shifts of  reflection light in the cold cavity and hot cavity, respectively. If the state of the injected electron spin is $|\uparrow\rangle$, the left circular light $|L\rangle$ gets a reflection phase shift $\varphi_{h}$, and the right circular light $|R\rangle$ has a reflection phase shift $\varphi_{o}$. If the injected electron spin is in the state $|\downarrow\rangle$, the right circular light $|R\rangle$ has a reflection phase shift $\varphi_{h}$, and the left circular light $|L\rangle$ gets a reflection phase shift $\varphi_{o}$. By adjusting $\omega$ and $\omega_{c}$ ($\omega-\omega_{c}=0$), $\varphi_{h}-\varphi_{o}=\pi$ can be obtained. In the condition $\omega=\omega_{X^{-}}=\omega_{c}$ and $\kappa\gg\kappa_s$, the reflection coefficients of cold and hot cavities can achieve $|r_{o}(\omega)|\simeq1$ and $|r_{h}(\omega)|\simeq1$. Then, the reflection operator $\hat{r}(\omega)$ can be expressed as
\begin{eqnarray}       
\begin{split}
\hat{r}(\omega)\;\;=\;\;& e^{i\varphi_{o}}[(|R\rangle\langle
R|\otimes|\uparrow\rangle\langle\uparrow|+|L\rangle\langle
L|\otimes|\downarrow\rangle\langle\downarrow|)\\
& +e^{i\varphi_{h}}(|L\rangle\langle
L|\otimes|\uparrow\rangle\langle\uparrow|+|R\rangle\langle
R|\otimes|\downarrow\rangle\langle\downarrow|)].
\end{split}
\end{eqnarray}
After performing reflection operator $\hat{r}(\omega)$ on the states of the photon and the electron spin, the input-output relation of the single-sided QD-cavity system can be obtained as
\begin{eqnarray}    
\begin{split}
&|R,\uparrow\rangle\;\;\rightarrow\;\;|r_{o}||R,\uparrow\rangle,\;\;\;\;\;\;\;\;\;\;\;
|L,\uparrow\rangle\;\;\rightarrow\;\;-|r_{h}||L,\uparrow\rangle, \\
&|R,\downarrow\rangle\;\;\rightarrow\;\;-|r_{h}||R,\downarrow\rangle,\;\;\;\;\;\;\;\;
|L,\downarrow\rangle\;\;\rightarrow\;\;|r_{o}||L,\downarrow\rangle.
\end{split}
\end{eqnarray}

\begin{center}
\begin{figure}[!h]
\includegraphics[width=4.2 cm,angle=0]{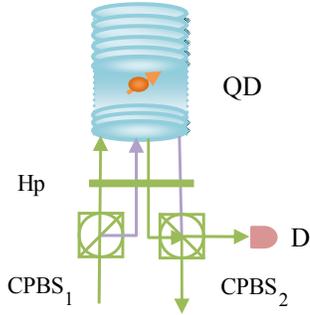}
\caption{ Schematic diagram of the error-heralded block. If the photon having passed through the block has not changed its own polarization state nor the state of the QD, it will be detected by the detector. Hp represents a half-wave plate, which performs the polarization Hadamard operation [$|R\rangle\rightarrow\frac{1}{\sqrt{2}}(|R\rangle+|L\rangle),|L\rangle\rightarrow\frac{1}{\sqrt{2}}(|R\rangle-|L\rangle)$
] on the photon. CPBS represents a polarizing beam splitter which transmits right circularly polarized photon $|R\rangle$ and reflects left circularly polarized photon $|L\rangle$. D represents a single-photon detector.}
\label{figure2}
\end{figure}
\end{center}

\section{The error-heralded block for the imperfect interaction between photons and a QD-cavity system}\label{sec3}

In order to avoid the errors caused by the imperfect interactions between photons and QD-cavity systems, we construct an error-heralded block with an QD in single-sided cavity system, two circular polarization beam splitters (CPBS), one half-wave plate (Hp) and one single-photon detector (D). The schematic diagram of the error-heralded block is shown in Fig.\;\ref{figure2}. The QD in the single-sided cavity is initially prepared in the state $\left| {{\varphi ^ + }} \right\rangle {\rm{ = }}\frac{1}{{\sqrt 2 }}(\left|  \uparrow  \right\rangle  + \left|  \downarrow  \right\rangle )$.

If the input photon is in the left-circularly polarized state $\left| L \right\rangle $, the state of the whole system composed of the photon and the QD in the cavity is
\begin{eqnarray}    
\left| {\Phi} \right\rangle_{0}\;\; =\;\;\left| L \right\rangle  \otimes \frac{1}{{\sqrt 2 }}(\left|  \uparrow  \right\rangle  + \left|  \downarrow  \right\rangle ).
\end{eqnarray}
The photon will be transmitted by the  CPBS on the left and then pass through the half-wave plate(Hp). The state of the whole system will change to $\left| {\Phi} \right\rangle_{1}$, which can be expressed as
\begin{eqnarray}    
\left| {\Phi} \right\rangle_{1}\;\;=\;\;\frac{1}{{\sqrt 2 }}(\left| R \right\rangle  - \left| L \right\rangle ) \otimes \frac{1}{{\sqrt 2 }}(\left|  \uparrow  \right\rangle  + \left|  \downarrow  \right\rangle ).
\end{eqnarray}
After the interaction between the photon and the QD-cavity system, the state becomes
\begin{small}
\begin{eqnarray}    
\left| {\Phi} \right\rangle_{2}=\frac{1}{2}({r_h}\left| R \right\rangle \left|  \uparrow  \right\rangle  + {r_o}\left| R \right\rangle \left|  \downarrow  \right\rangle  - {r_o}\left| L \right\rangle \left|  \uparrow  \right\rangle  - {r_h}\left| L \right\rangle \left|  \downarrow  \right\rangle ).
\end{eqnarray}
\end{small}
Then the photon passes through the half-wave plate(Hp) again, the state will change to
\begin{eqnarray}   
\begin{split}
\left| {\Phi} \right\rangle_{3}&=
\frac{1}{{2\sqrt 2 }}[{r_h}(\left| R \right\rangle+\left| L \right\rangle )\left|  \uparrow  \right\rangle+{r_o}(\left| R \right\rangle+\left| L \right\rangle )\left|  \downarrow  \right\rangle\\
&-{r_o}(\left| R \right\rangle-\left| L \right\rangle )\left|  \uparrow  \right\rangle  -{r_h}(\left| R \right\rangle-\left| L \right\rangle )\left|  \downarrow  \right\rangle ]\\
 &=\frac{1}{{2\sqrt 2 }}[({r_h}-{r_o})\left| R \right\rangle \left|  \uparrow  \right\rangle+({r_h}+{r_o})\left| L \right\rangle \left|  \uparrow  \right\rangle\\
 &-({r_h}-{r_o})\left| R \right\rangle \left|  \downarrow  \right\rangle-({r_h}+{r_o})\left| L \right\rangle \left|  \downarrow  \right\rangle ]\\
 &=\frac{1}{{2\sqrt 2 }}[({r_h}-{r_o})\left| R \right\rangle (\left|  \uparrow  \right\rangle-\left|  \downarrow  \right\rangle )\\
 &+({r_h}+ {r_o})\left| L \right\rangle (\left|  \uparrow  \right\rangle+\left|  \downarrow  \right\rangle )]\\
 &=\frac{1}{2}[({r_h}-{r_o})\left| R \right\rangle \left| {{\varphi^{-}}} \right\rangle+({r_h}+{r_o})\left| L \right\rangle \left| {{\varphi ^+}} \right\rangle ].
 \end{split}
\end{eqnarray}

At last, the photon passes through the CPBS on the right and the reflected $\left| L \right\rangle $ component will be detected by the single photon detector (D), and the final state can be described as
\begin{eqnarray}    
\left| {\Phi} \right\rangle_{4}\;\;=\;\;\frac{1}{2}({r_h} - {r_o})\left| R \right\rangle \left| {{\varphi ^ - }} \right\rangle.
\end{eqnarray}
We can see that if the photon is reflected by the error-heralded block, the polarization of the photon and the state of the QD would not change, the photon will be detected by the detector and the click of detector represents the error-herald process. If there is no click of the detector, the photon will transmitted from the
error-heralded block and means there are no errors in the process.

\section{Error-heralded generation of hyperentangled Bell states
for two photons interacting with single-sided quantum-dot-cavity systems}\label{sec4}

\begin{center}
\begin{figure}[!h]
\includegraphics[width=7.2 cm,angle=0]{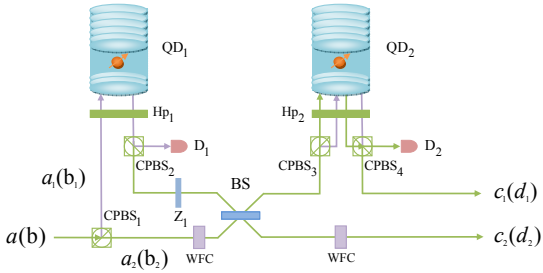}
\caption{Schematic diagram of the error-heralded two-photon HBSG in both polarization and spatial modes. BS is a 50 : 50 beam splitter which performs the spatial-mode Hadamard operation [$|x_{1}\rangle\rightarrow\frac{1}{\sqrt{2}}(|y_{1}\rangle+|y_{2}\rangle),
|x_{2}\rangle\rightarrow\frac{1}{\sqrt{2}}(|y_{1}\rangle-|y_{2}\rangle)$,
$x=a,b$ and $y=c,d$]. $Z_{1}$ is a half-wave plate which performs a polarization bit-flip operation
$Z=|R\rangle \langle L|+|L\rangle \langle R|$. WFC is a waveform corrector mapping $|i\rangle_{2}$ to $\frac{1}{2}(r_{h}-r_{o})|i\rangle_{2}$, $|i\rangle_{1}$ and $|i\rangle_{2}$ denote two spatial modes of photon $i$ ($i=a,b$).} \label{figure3}
\end{figure}
\end{center}

 The hyperentangled Bell states in both polarization and spatial-mode DOFs can be written as
\begin{eqnarray}    
|\Phi\rangle_{AB}\;\;=\;\; |\varphi_{P}\rangle_{AB}\otimes|\varphi_{S}\rangle_{AB}.
\end{eqnarray}
Here $ AB $ denotes two photons. $|\varphi_{P}\rangle_{AB}$ is one of the four polarization Bell state:
\begin{eqnarray}  
\begin{split}
|\phi^{\pm}_{P}\rangle_{AB}\;\;=\;\;\frac{1}{\sqrt{2}}(|RR\rangle\pm|LL\rangle),\\
|\psi^{\pm}_{P}\rangle_{AB}\;\;=\;\;\frac{1}{\sqrt{2}}(|RL\rangle\pm|LR\rangle).
\end{split}
\end{eqnarray}
$|\varphi_{S}\rangle_{AB}$ is one of the four spatial-mode Bell states:
\begin{eqnarray}  
\begin{split}
|\phi^{\pm}_{S}\rangle_{AB}\;\;=\;\;\frac{1}{\sqrt{2}}(|a_{1}b_{1}\rangle\pm|a_{2}b_{2}\rangle),\\
|\psi^{\pm}_{S}\rangle_{AB}\;\;=\;\;\frac{1}{\sqrt{2}}(|a_{1}b_{2}\rangle\pm|a_{2}b_{1}\rangle).
\end{split}
\end{eqnarray}
Here $a_1$ ($b_1$) and $a_2$ ($b_2$) denote the two spatial modes of photon $A$ ($B$). The states $|\phi^{\pm}_{P}\rangle_{AB}$ and $|\phi^{\pm}_{S}\rangle_{AB}$ are in even-parity mode, and the states $|\psi^{\pm}_{P}\rangle_{AB}$ and $|\psi^{\pm}_{S}\rangle_{AB}$ are in odd-parity mode.
The principle for the hyperentangled Bell states generation (HBSG) is shown in Fig.\;\ref{figure3}. The initial states of the two electron spins e$_1$ and e$_2$ in two QD-cavity systems QD$_1$ and QD$_2$ are $|\varphi^{+}\rangle_1$ and $|\varphi^{+}\rangle_2$, respectively. Here $|\varphi^{\pm}\rangle=\frac{1}{\sqrt{2}}(|\uparrow\rangle\pm|\downarrow\rangle)$, and the two QD-cavity systems QD$_1$ and QD$_2$ are in the ideal condition. The two photons A and B are prepared in the same initial state $|\varphi^{+}\rangle_A=|\varphi^{+}\rangle_B=\frac{1}{\sqrt{2}}(|R\rangle+|L\rangle)$, the process for HBSG can be described in detail as follows.

First, photon A is injected into the quantum circuit through the left input port, then enters photon B. It must be ensured that the time interval between the two input photons is less than the decoherence time of the electrons in QD. CPBS$_{1}$ will transmit the photons in state $|R\rangle$ to path $a_{2}(b_{2})$ and reflect the $|L\rangle$ photons to path $a_{1}(b_{1})$. The photons in path $a_{1}(b_{1})$ will then pass through the error-detected block QD$_{1}$ and Hp$_{1}$. The photons in path $a_{2}(b_{2})$ will then pass through the WFC (waveform corrector) \cite{Li1}, which changes $|i\rangle_{2}$ to $\frac{1}{2}(r_{h}-r_{o})|i\rangle_{2}$. This slightly
decreases the overall success probability, but leaves the
fidelity intact. The state of the entire system changes from $\Psi_{0}=|\varphi^{+}\rangle_A|\varphi^{+}\rangle_B|\varphi^{+}\rangle_1|\varphi^{+}\rangle_2$ to $\Psi_{1}$,  before the photons reach the BS. Here $\Psi_{1}$ is
\begin{eqnarray}   
\begin{split}
\Psi_{1}\;\;=\;\;&\frac{1}{2}(|LLa_{1}(b_{1}\rangle|\varphi^{+}\rangle_1+|LRa_{1}(b_{2}\rangle|\varphi^{-}\rangle_1\\
&+|RLa_{2}(b_{1}\rangle|\varphi^{-}\rangle_1+|RRa_{2}(b_{2}\rangle|\varphi^{+}\rangle_1)|\varphi^{+}\rangle_2.
\end{split}
\end{eqnarray}
Then the two wave packets splitted by CPBS$_{1}$ will interfere at BS, which complete the spatial-mode Hadamard operation [$|a_{1}(b_{1})\rangle\rightarrow\frac{1}{\sqrt{2}}(|c_{1}(d_{1})\rangle+|c_{2}(d_{2})\rangle),
|a_{2}(b_{2})\rangle\rightarrow\frac{1}{\sqrt{2}}(|c_{1}(d_{1})\rangle-|c_{2}(d_{2})\rangle)]$, and lead photons to path $c_{1}(d_{1})$ or $c_{2}(d_{2})$. Photons in path $c_{1}(d_{1})$ will then pass through the second error-heralded block consisting of the two CPBSs, QD$_{2}$ and Hp$_{2}$. The whole system will change to the following state
\begin{small}
\begin{eqnarray}    
\begin{split}
\Psi_{2}=&\frac{1}{2}(|\phi^{+}\rangle_P|\phi^{+}\rangle_S|\varphi^{+}\rangle_1|\varphi^{+}\rangle_2
+|\psi^{-}\rangle_P|\psi^{-}\rangle_S|\varphi^{+}\rangle_1|\varphi^{-}\rangle_2\\
&+|\psi^{+}\rangle_P|\phi^{-}\rangle_S|\varphi^{-}\rangle_1|\varphi^{+}\rangle_2
-|\phi^{-}\rangle_P|\psi^{+}\rangle_S|\varphi^{-}\rangle_1|\varphi^{-}\rangle_2).
\end{split}
\end{eqnarray}
\end{small}
 Eq.(16) shows the relationship between the polarization-spatial hyperentangled Bell states of the two photons syetem and the measurement results of the two QD-cavity systems. If QD$_{1}$ and QD$_{2}$ are in the states $|\varphi^{+}\rangle_1$ and $|\varphi^{+}\rangle_2$, respectively, the two photon is in the hyperentangled Bell state $|\phi^{+}\rangle_P|\phi^{+}\rangle_S$. When QD$_{1}$ and QD$_{2}$ are in the states $|\varphi^{+}\rangle_1$ and $|\varphi^{-}\rangle_2$, respectively, the two photon is in the hyperentangled Bell state $|\psi^{-}\rangle_P|\psi^{-}\rangle_S$. Similarly, QD states $|\varphi^{-}\rangle_1|\varphi^{+}\rangle_2$ corresponds to hyperentangled Bell state $|\psi^{+}\rangle_P|\phi^{-}\rangle_S$, QD states $|\varphi^{-}\rangle_1|\varphi^{-}\rangle_2$ corresponds to hyperentangled Bell state $|\phi^{-}\rangle_P|\psi^{+}\rangle_S$. In this way, we can deterministically generate 4 two-photon polarization-spatial hyperentangled Bell states by measuring the state of two QDs, and then generate the other 12 polarization-spatial hyperentangled Bell states through some appropriate single-bit operations.

\begin{center}
\begin{figure}[!h]
\includegraphics[width=7.2 cm,angle=0]{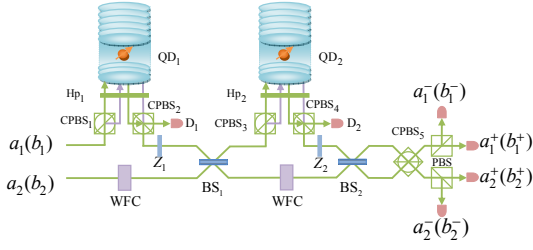}
\caption{Schematic diagram of the self-assisted two-photon complete polarization-spatial HBSA. } \label{HBSA}
\end{figure}
\end{center}

\section{Self-assisted complete two-photon polarization-spatial-mode hyperentangled Bell states analysis with error-heralded blocks}\label{sec5}

 The principle for distinguishing the 16 hyperentangled Bell states is shown in Fig.\;\ref{HBSA}. Which contains two steps: spatial-mode Bell states analysis and polarization Bell states analysis. The two error-heralded block are used to record the space-state information. The Z represents a half-wave plate which is used to perform a bit-flip operation $Z=|R\rangle \langle L|+|L\rangle \langle R|$ in
the polarization DOF, and the beam splitters(BSs) guide the photons from each input port to those two output ports with equal probabilities. The waveform correctors(WFCs) in path $a_{2}(b_{2})$ is used to eliminate the infidelity caused by the cavity-QD-system. The circular polarization beam splitter(CPBS) transmits the input right-circularly polarized photon and reflects the left-circularly polarized photon, the polarization beam splitter(PBS) transmit horizontal polarized states while reflecting vertical polarized states. The CPBS and two PBSs form a single-photon Bell state measurement device (SPBSM), which discriminates four single-photon Bell states completely, assisted by the spatial information recorded in the two QDs, we can completely distinguish the 16 two-photon hyperentangled Bell states. We describe the process for HBSA as follows.

\subsection{HBSA protocol for Bell states in spatial mode}\label{}

The first step is used to distinguish the 4 spatial-mode Bell states assisted by the single-sided QD-cavity systems. The initial states of the two electron spins e$_1$ and e$_2$ in two QD-cavity systems QD$_1$ and QD$_2$ are $|\varphi^{+}\rangle_1=|\varphi^{+}\rangle_2$, and the two QD-cavity systems QD$_1$ and QD$_2$ are in the ideal condition. The state of photon pair $AB$ is one of the 16 hyperentangled Bell states. Then let photons A and B enter the state analysis device shown in Fig. \ref{HBSA}, one after another, the interval time between the two photons should be less than the spin coherence time of the QDs. After the two photons interacting with QD-cavity systems QD$_1$, before the two wave packets in
mode $a(b)_{1}$ and $a(b)_{2}$ interfere at BS$_{1}$, the state of the system composed of electron spin e$_1$ and photon pair $AB$ will changed to:
\begin{eqnarray}    
\begin{split}
&|\phi^{\pm}_{P}\rangle_{AB}|\phi^{+}_{S}\rangle_{AB}\;\xrightarrow[a_{1},b_{1}]{QD_1,Z_{1}}\;  |\phi^{\pm}_{P}\rangle_{AB}|\phi^{+}_{S}\rangle_{AB}|\varphi^{+}\rangle_1, \\
&|\psi^{\pm}_{P}\rangle_{AB}|\phi^{+}_{S}\rangle_{AB}\;\xrightarrow[a_{1},b_{1}]{QD_1,Z_{1}}\;  |\psi^{\pm}_{P}\rangle_{AB}|\phi^{+}_{S}\rangle_{AB}|\varphi^{+}\rangle_1, \\
&|\phi^{\pm}_{P}\rangle_{AB}|\phi^{-}_{S}\rangle_{AB}\;\xrightarrow[a_{1},b_{1}]{QD_1,Z_{1}}\;  |\phi^{\pm}_{P}\rangle_{AB}|\phi^{-}_{S}\rangle_{AB}|\varphi^{+}\rangle_1, \\
&|\psi^{\pm}_{P}\rangle_{AB}|\phi^{-}_{S}\rangle_{AB}\;\xrightarrow[a_{1},b_{1}]{QD_1,Z_{1}}\;  |\psi^{\pm}_{P}\rangle_{AB}|\phi^{-}_{S}\rangle_{AB}|\varphi^{+}\rangle_1, \\
&|\phi^{\pm}_{P}\rangle_{AB}|\psi^{+}_{S}\rangle_{AB}\;\xrightarrow[a_{1},b_{1}]{QD_1,Z_{1}}\;  |\phi^{\pm}_{P}\rangle_{AB}|\psi^{+}_{S}\rangle_{AB}|\varphi^{-}\rangle_1, \\
&|\psi^{\pm}_{P}\rangle_{AB}|\psi^{+}_{S}\rangle_{AB}\;\xrightarrow[a_{1},b_{1}]{QD_1,Z_{1}}\;  |\psi^{\pm}_{P}\rangle_{AB}|\psi^{+}_{S}\rangle_{AB}|\varphi^{-}\rangle_1, \\
&|\phi^{\pm}_{P}\rangle_{AB}|\psi^{-}_{S}\rangle_{AB}\;\xrightarrow[a_{1},b_{1}]{QD_1,Z_{1}}\;  |\phi^{\pm}_{P}\rangle_{AB}|\psi^{-}_{S}\rangle_{AB}|\varphi^{-}\rangle_1, \\
&|\psi^{\pm}_{P}\rangle_{AB}|\psi^{-}_{S}\rangle_{AB}\;\xrightarrow[a_{1},b_{1}]{QD_1,Z_{1}}\;  |\psi^{\pm}_{P}\rangle_{AB}|\psi^{-}_{S}\rangle_{AB}|\varphi^{-}\rangle_1.
\end{split}
\end{eqnarray}
From the above changes, it can be seen that QD$_1$ records the parity information of the spatial-mode state. If the electron spin e$_1$ is in the state $|\varphi^{+}\rangle_1$, the spatial-mode state of photon pair $AB$ is in the even-parity mode. If the electron spin e$_1$ is in the state $|\varphi^{-}\rangle_1$, the spatial-mode state of photon pair $AB$ is in the odd-parity mode. Now, we can see that the even-parity Bell states can be distinguish from the odd-parity Bell states assisted by the QD-cavity systems QD$_1$, without affecting the state of the photon pair $AB$.
Then the photons in two spatial-mode will interfere at BS$_{1}$, which performs a Hadamard operation on the spatial-mode state while transforming the phase information into parity information. Therefore we can similarly use QD$_2$ to record the phase information of the spatial-mode state. The corresponding transformations on the states before the c-PBS can be described as follows:
\begin{small}
\begin{eqnarray}   
\begin{split}
&|\phi^{\pm}_{P}\rangle_{AB}|\phi^{+}_{S}\rangle_{AB}\xrightarrow[a_{1},b_{1}]{QD_{1},Z_{1},QD_{2},Z_{2}}\;  |\phi^{\pm}_{P}\rangle_{AB}|\phi^{+}_{S}\rangle_{AB}|\varphi^{+}\rangle_{1}|\varphi^{+}\rangle_{2}, \\
&|\psi^{\pm}_{P}\rangle_{AB}|\phi^{+}_{S}\rangle_{AB}\xrightarrow[a_{1},b_{1}]{QD_{1},Z_{1},QD_{2},Z_{2}}\; |\psi^{\pm}_{P}\rangle_{AB}|\phi^{+}_{S}\rangle_{AB}|\varphi^{+}\rangle_{1}|\varphi^{+}\rangle_{2}, \\
&|\phi^{\pm}_{P}\rangle_{AB}|\phi^{-}_{S}\rangle_{AB}\xrightarrow[a_{1},b_{1}]{QD_{1},Z_{1},QD_{2},Z_{2}}\; |\phi^{\pm}_{P}\rangle_{AB}|\phi^{-}_{S}\rangle_{AB}|\varphi^{+}\rangle_{1}|\varphi^{-}\rangle_{2}, \\
&|\psi^{\pm}_{P}\rangle_{AB}|\phi^{-}_{S}\rangle_{AB}\xrightarrow[a_{1},b_{1}]{QD_{1},Z_{1},QD_{2},Z_{2}}\; |\psi^{\pm}_{P}\rangle_{AB}|\phi^{-}_{S}\rangle_{AB}|\varphi^{+}\rangle_{1}|\varphi^{-}\rangle_{2}, \\
&|\phi^{\pm}_{P}\rangle_{AB}|\psi^{+}_{S}\rangle_{AB}\xrightarrow[a_{1},b_{1}]{QD_{1},Z_{1},QD_{2},Z_{2}}\; |\phi^{\pm}_{P}\rangle_{AB}|\psi^{+}_{S}\rangle_{AB}|\varphi^{-}\rangle_{1}|\varphi^{+}\rangle_{2}, \\
&|\psi^{\pm}_{P}\rangle_{AB}|\psi^{+}_{S}\rangle_{AB}\xrightarrow[a_{1},b_{1}]{QD_{1},Z_{1},QD_{2},Z_{2}}\; |\psi^{\pm}_{P}\rangle_{AB}|\psi^{+}_{S}\rangle_{AB}|\varphi^{-}\rangle_{1}|\varphi^{+}\rangle_{2}, \\
&|\phi^{\pm}_{P}\rangle_{AB}|\psi^{-}_{S}\rangle_{AB}\xrightarrow[a_{1},b_{1}]{QD_{1},Z_{1},QD_{2},Z_{2}}\;  |\phi^{\pm}_{P}\rangle_{AB}|\psi^{-}_{S}\rangle_{AB}|\varphi^{-}\rangle_{1}|\varphi^{-}\rangle_{2}, \\
&|\psi^{\pm}_{P}\rangle_{AB}|\psi^{-}_{S}\rangle_{AB}\xrightarrow[a_{1},b_{1}]{QD_{1},Z_{1},QD_{2},Z_{2}}\; |\psi^{\pm}_{P}\rangle_{AB}|\psi^{-}_{S}\rangle_{AB}|\varphi^{-}\rangle_{1}|\varphi^{-}\rangle_{2}.
\end{split}
\end{eqnarray}
\end{small}
After passing through the two QD systems, the two photons A and B can be entangled with the electron spins in the two cavities.
It can be seen that such a two-photon hyperentangled state does not change after interacting with the QDs, Hps, BSs, the whole system returns to the initial state. Different combinations of two QDs status, reflecting four different two-photon entanglement spatial states.
We assume that the initial states of the excess electron in the cavity is
$|\varphi^{+}\rangle=\frac{1}{\sqrt{2}}(|\uparrow\rangle+|\downarrow\rangle)$. The Z represents a half-wave plate which is used to perform a bit-flip operation $Z=|R\rangle \langle L|+|L\rangle \langle R|$ in the polarization DOF, Hp represents a half-wave plate, which can performs the polarization Hadamard operation $|R\rangle\rightarrow\frac{1}{\sqrt{2}}(|R\rangle+|L\rangle),|L\rangle\rightarrow\frac{1}{\sqrt{2}}(|R\rangle-|L\rangle)$
on the photon, and the beam splitters(BSs) can accomplish the Hadamard operation on the spatial-mode DOF. The circular polarization beam splitter(c-PBS) transmits the input right-circularly polarized photon and reflects the left-circularly polarized photon, the polarization beam splitter(PBS) transmit horizontal polarized states while reflecting vertical polarized states. The two QDs are used to record spatial information and do not destroy the polarization information, QD$_{1}$ is used to record the parity information, and QD$_{2}$ to record the phase information. The spin state of excess electron in QD$_{1}$ is changed for odd-parity states and unchanged for even-parity. The spin state of excess electron in QD$_{2}$ is changed for negative phase states and unchanged for positive phase states. The relation between the outcomes of the two QDs and the initial space states is shown in Table I.

We can detect whether or not there is a photon interacting with the QD-cavity system by measuring the spin state of the excess electron. If the excess electron is in the initial states, there is no photon (or two photons) interacting with the QD-cavity system. If the excess electron is in the $|\varphi^{-}\rangle=\frac{1}{\sqrt{2}}(|\uparrow\rangle-|\downarrow\rangle)$ states, there is a photon interacting with the QD cavity system (with a bit-flip operation $Z=|R\rangle \langle L|+|L\rangle \langle R|$ on the photon, its original polarization state is recovered). At this point, we can fully distinguish the four space Bell states and do not affect it.

\begin{table}[!h]
\centering \caption{The relation between the outcomes of the two QDs
 and the initial space states.}
\begin{tabular}{ccc}
\hline\hline
  Initial space states    & QD$_1$                 &     $\;\;\;\;\;\;\;\;$ QD$_2$ $\;\;\;\;\;\;\;\;$  \\
   \hline
$|\phi_{S}^{+}\rangle_{AB}$&$|\varphi^{+}\rangle_1$   &       $|\varphi^{+}\rangle_2$ \\
$|\phi_{S}^{-}\rangle_{AB}$&$|\varphi^{+}\rangle_1$   &       $|\varphi^{-}\rangle_2$ \\
 $|\psi_{S}^{+}\rangle_{AB}$&$|\varphi^{-}\rangle_1$   &       $|\varphi^{+}\rangle_2$ \\
$|\psi_{S}^{-}\rangle_{AB}$ &$|\varphi^{-}\rangle_1$   &       $|\varphi^{-}\rangle_2$  \\
\hline\hline
\end{tabular}\label{table1}
\end{table}

\subsection{HBSA protocol for Bell states in polarization mode}\label{}

The second step is discrimination of 4 polarization Bell states assisted by the spatial-mode entanglement. The C-PBS and two PBSs form a single-photon Bell state measurement device (SPBSM), which can be used to discriminate 4 single-photon polarization Bell states completely. The 8 single-photon Bell states of two photons $A$,
$B$ that composed of the polarization and spatial-mode DOFs
can be expressed as
\begin{eqnarray}    
\begin{split}
|\phi^{\pm}\rangle_{X}=\frac{1}{\sqrt{2}}(|Rx_{2}\rangle\pm|Lx_{1}\rangle),\\
|\psi^{\pm}\rangle_{X}=\frac{1}{\sqrt{2}}(|Rx_{1}\rangle\pm|Lx_{2}\rangle).
\end{split}
\end{eqnarray}
Here X(x) can be either A(a) or B(b). For example, after the photon $A$ passes through c-PBS, the four single-photon Bell states will change to
\begin{eqnarray}   
\begin{split}
|\phi^{+}\rangle_{A}\rightarrow\frac{1}{\sqrt{2}}(|R\rangle+|L\rangle)|a_{1}\rangle=|H\rangle|a_{1}\rangle,\\
|\phi^{-}\rangle_{A}\rightarrow\frac{1}{\sqrt{2}}(|R\rangle-|L\rangle)|a_{1}\rangle=|V\rangle|a_{1}\rangle,\\
|\psi^{+}\rangle_{A}\rightarrow\frac{1}{\sqrt{2}}(|R\rangle+|L\rangle)|a_{2}\rangle=|H\rangle|a_{2}\rangle,\\
|\psi^{-}\rangle_{A}\rightarrow\frac{1}{\sqrt{2}}(|R\rangle-|L\rangle)|a_{2}\rangle=|V\rangle|a_{2}\rangle.
\end{split}
\end{eqnarray}
Here $|R\rangle=\frac{1}{\sqrt{2}}(|H\rangle+|V\rangle)$ and $|L\rangle=\frac{1}{\sqrt{2}}(|H\rangle-|V\rangle)$. After the photon $A$ passing through two PBSs, the photon in state $|H\rangle|a_{1}\rangle$, $|V\rangle|a_{1}\rangle$, $|H\rangle|a_{2}\rangle$ or $|V\rangle|a_{2}\rangle$ will be detected by the photon detector $a^{+}_{1}$, $a^{-}_{1}$, $a^{+}_{2}$ or $a^{-}_{2}$, respectively. The four single-photon Bell states of photon $B$ can be detected in the same way. The relationship of the 8 single-photon Bell states and the corresponding response of photon detectors is summarized as follows:
\begin{eqnarray}    
|\phi^{\pm}\rangle_{X}\;\longleftrightarrow\; x^{\pm}_{1},\;\;\;\;\;\;
|\psi^{\pm}\rangle_{X}\;\longleftrightarrow\; x^{\pm}_{2}.
\end{eqnarray}
Here X(x) can be either A(a) or B(b). Before performing the single-photon Bell state measurements(SPBSMs), the spatial-mode state of the two photons is known, and can assist the analysis of the four polarization Bell states. If the spatial-mode state after the first step is $|\phi_{S}^{+}\rangle_{AB}$, the four possible combinations of SPBSMs will be
\begin{eqnarray}    
\begin{split}
|\phi_{P}^{\pm}\rangle_{AB}|\phi_{S}^{+}\rangle_{AB}
\;\;=\;\;&\frac{1}{2}(|\phi^{\pm}\rangle_{A}\phi^{+}\rangle_{B}+|\phi^{\mp}\rangle_{A}\phi^{-}\rangle_{B}\\
&+|\psi^{\pm}\rangle_{A}\psi^{+}\rangle_{B}+|\psi^{\mp}\rangle_{A}\psi^{-}\rangle_{B}),\\
|\psi_{P}^{\pm}\rangle_{AB}|\phi_{S}^{+}\rangle_{AB}
=\;\;&\frac{1}{2}(|\phi^{\pm}\rangle_{A}\psi^{+}\rangle_{B}-|\phi^{\mp}\rangle_{A}\psi^{-}\rangle_{B}\\
&+|\psi^{\pm}\rangle_{A}\phi^{+}\rangle_{B}-|\psi^{\mp}\rangle_{A}\phi^{-}\rangle_{B}).
\end{split}
\end{eqnarray}
There are 16 possible measurement combinations, which can be divided into four different groups, each group consists of 4 different polarization and space Bell states. The relationship between the initial hyperentangled states and the possible detections is shown in Table II. We can determine which group the probe result belongs to, assisted by the known spatial information, it is possible to completely distinguish the 4 polarization states. And finally determine the original hyper-entangled Bell states of the two-photon.

\begin{table}[htb]
\centering
\caption{Relationship between the initial hyperentangled states and the possible detections.}
\begin{tabular}{cc}
\hline\hline
 $\;\;\; $   Initial states    $\;\;\;$  &      $\;\;\;\;\;\;\;\;\;$ possible detections  $\;\;\; $        \\
   \hline
$|\phi^{+}_{P}\rangle|\phi^{+}_{S}\rangle,|\psi^{+}_{P}\rangle|\psi^{+}_{S}\rangle$
&$\;\;\;\;$  $a^{+}_{1}b^{+}_{1},a^{+}_{2}b^{+}_{2}$ \\
$|\phi^{-}_{P}\rangle|\phi^{-}_{S}\rangle,|\psi^{-}_{P}\rangle|\psi^{-}_{S}\rangle$
&$\;\;\;\;$ $a^{-}_{1}b^{-}_{1},a^{-}_{2}b^{-}_{2}$\\
$|\phi^{+}_{P}\rangle|\phi^{+}_{S}\rangle,|\phi^{+}_{P}\rangle|\phi^{+}_{S}\rangle$
&$\;\;\;\;$  $a^{+}_{1}b^{+}_{2},a^{+}_{2}b^{+}_{1}$ \\
$|\phi^{-}_{P}\rangle|\psi^{-}_{S}\rangle,|\psi^{-}_{P}\rangle|\phi^{-}_{S}\rangle$
&$\;\;\;\;$ $a^{-}_{1}b^{-}_{2},a^{-}_{2}b^{-}_{1}$\\
$|\phi^{+}_{P}\rangle|\phi^{-}_{S}\rangle,|\psi^{+}_{P}\rangle|\psi^{-}_{S}\rangle$
&$\;\;\;\;$  $a^{+}_{1}b^{-}_{1},a^{+}_{2}b^{-}_{2}$ \\
$|\phi^{-}_{P}\rangle|\phi^{+}_{S}\rangle,|\psi^{-}_{P}\rangle|\psi^{+}_{S}\rangle$
&$\;\;\;\;$ $a^{-}_{1}b^{+}_{1},a^{-}_{2}b^{+}_{2}$\\
$|\phi^{+}_{P}\rangle|\psi^{-}_{S}\rangle,|\psi^{+}_{P}\rangle|\phi^{-}_{S}\rangle$
&$\;\;\;\;$  $a^{+}_{1}b^{-}_{2},a^{+}_{2}b^{-}_{1}$ \\
$|\phi^{-}_{P}\rangle|\psi^{+}_{S}\rangle,|\psi^{-}_{P}\rangle|\phi^{+}_{S}\rangle$
&$\;\;\;\;$ $a^{-}_{1}b^{+}_{2},a^{-}_{2}b^{+}_{1}$\\
\hline\hline
\end{tabular}\label{table2}
\end{table}

For example if the two QDs are both unchanged, their states are $|\varphi^{+}\rangle_1|\varphi^{+}\rangle_2$, it means that the space state is $|\phi_{S}^{+}\rangle$, and if the detector $a^{+}_{1}$ and $b^{+}_{1}$ have a response, it means that the single photon Bell state is $|\phi^{+}\rangle_{A}|\phi^{+}\rangle_{B}$, from table II we can see it belongs to the first group , so the polarization state is $|\phi_{P}^{+}\rangle$, the original hyperentangled Bell states of the two photons is $|\phi_{P}^{+}\rangle|\phi_{S}^{+}\rangle$. If the detector $a^{-}_{1}$ and $b^{+}_{2}$ clicked, it means that the single photon Bell state is $|\phi^{-}\rangle_{A}|\psi^{+}\rangle_{B}$, from table II we can see it belongs to the fourth group , and we also know the space state is $|\phi_{S}^{+}\rangle$, so the polarization state is $|\psi_{P}^{-}\rangle$, the original hyper-entangled Bell states of the two-photon is $|\psi_{P}^{-}\rangle|\phi_{S}^{+}\rangle$.  Similarly, we can completely distinguish 16 two-photon hyperentangled Bell states with the state of two QDs and the response of eight single photon detectors.

\section{Discussion and conclusion}\label{sec6}

Since the deterministic HBSG and the complete HBSA of the two-photon hyperentangled state are very important in quantum information process, a considerable number of theoretical and experimental programs have been proposed \cite{hyper1,hyper2,hyper3,hyper4,hyper5,hyper6,Sheng2,Li,Liu1,Wang1,WangGYHBSA,Ren,LiThereldedrepeater}, each of which has both strengths and weakness. Schemes involving only linear optical element cannot deterministically generate hyperentangled state nor completely  distinguish hyperentangled states. While those employing nonlinear materials such as QDs in microcavity can accomplish the deterministic HBSG and complete HBSA, due to the imperfect interactions of photons with quantum dot systems, the fidelities
and the efficiencies can hardly reach unit one. Based on the optical transitions in a QD-cavity system, we constructed an error-heraled block.
With this error-detected block, we proposed two self-assisted schemes for the deterministic HBSG and complete HBSA of two-photon polarization-spatial-mode hyperentangled system. The errors due to imperfect interactions between photons and quantum dot systems can be heralded by detectors, which makes the fidelities of the schemes reach unit one. In an ideal condition, $\left| {\Phi} \right\rangle_{f1}=\frac{1}{2}({r_h} - {r_o})\left| R \right\rangle \left| {{\varphi ^ - }} \right\rangle$ can be obtained after the left-polarized photon $\left| L \right\rangle$ interacting with the error-detected block and the fidelities and the efficiencies can be 100$\%$. However, in a realistic condition, the outcomes of the interaction between the photon and the error-heraled block, which are described as Eq. (10), are affected by the coupling sthength $g$ of the quantum dot and microcavity, the decay rates of $X^{-}$ and microcavity field mode $\gamma/2$ and $\kappa/2$, the microcavity leakage rate $\kappa_{s}/2$, which would affect the fidelities and the efficiencies as well.

\begin{center}
\begin{figure}[!h]
\includegraphics[width=6 cm,angle=0]{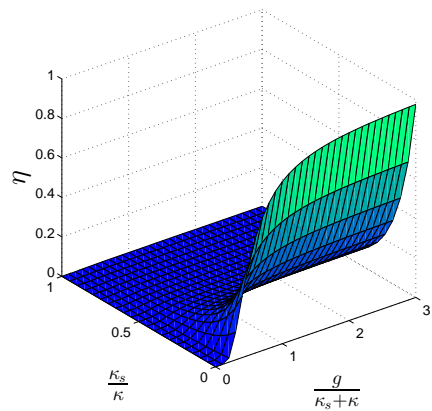}
\caption{The efficiency of the deterministic HBSG and complete HBSA  protocols vs $\frac{\kappa_{s}}{\kappa}$ and $\frac{g}{\kappa_{s}+\kappa}$, $\gamma=0.1\kappa$ under the condition $\omega=\omega_{X^{-}}=\omega_{c}$.}
\label{figure5}
\end{figure}
\end{center}

The fidelity of the process for deterministic HBSG and complete HBSA is
defined as $F=|\langle\psi_{f}|\psi\rangle|^{2}$, where $|\psi_{f}\rangle$ is the final state of the total system under actual circumstances and $|\psi\rangle$ is the final state with an ideal condition. Because the schemes add WFC(wave form corrector) to the path that has no nonlinear interaction, the fidelity can be kept in unit one. The efficiency is defined as the ratio of the number of the output photons to the number of the input photons. Since in the two schemes photons passed through the same number of nonlinear materials and the fidelity is one, the efficiency of the state generation scheme and the state analysis scheme are the same $\eta_{HBSG}=\eta_{HBSA}=\frac{1}{2^{8}}(r_{h}-r_{o})^{8}$. In Fig.\;\ref{figure5}, we numerically simulate the efficiency of the protocols.

Besides the parameters mentioned above, the exciton
dephasing, including the optical dephasing and the spin dephasing of $X^{-}$ will also affect the fidelities. Exciton dephasing reduces the fidelity by the amount of $(1-\exp^{-\tau/\Gamma})$, where $\tau$ and $ \Gamma$ are the cavity photon lifetime and the trion coherence time, respectively. The optical dephasing reduces the fidelity
less than 10$\%$, because the time scale of the excitons can reach hundreds of picoseconds \cite{Borri,Birkedal,Langbein}, while the cavity photon lifetime is in the tens of picoseconds range for a self-assembled
In(Ga)As-based QD with a cavity Q factor in the strong coupling regime. The effect of the spin dephasing can be neglected because the spin decoherence time is several orders of magnitude longer than the cavity photon lifetime  \cite{Heiss,Gerardot,Brunner}.

In summary, we have proposed two separate schemes for deterministic HBSG and self-assisted complete HBSA of the two-photon polarization-spatial-mode hyperentangled states with error-heralded blocks. With the help of the error-detected block, the errors can be heralded by detectors, which makes the fidelity of the protocols reach unit one. In the proposal, the four spatial-mode Bell states are completely distinguished by using two QD-cavity systems without affecting the hyperentangled state of the two-photon system, so the four polarization Bell states can be
completely distinguished only by linear optical elements assisted by the polarization-spatial hyperentanglement. The protocols greatly reduced the difficulty of experimental realization. We only need one QD for one photon, and the device is much simpler than previous, taking a step closer to future long-distance high capacity quantum communication.
\section*{Acknowledgment}

This work was supported by the National Natural Science Foundation of China (No. 11475021); National Key Basic Research Program of China (No. 2013CB922000).

\section*{Conflict of interests}
The authors declare that they have no conflict of interests.

\end{document}